\documentclass{elsart}

\usepackage{epsfig}

\usepackage{amssymb}

\begin{document}

\begin{frontmatter}



\title{Study of the generation intensity in the backward wave oscillator with a ''grid'' diffraction grating
as a function of the grating length}


\author{V.G. Baryshevsky},
\author{N.A.Belous,}
\author{V.A.Evdokimov,}
\author{A.A.Gurinovich,}
\author{A.S.Lobko,}
\author{P.V.Molchanov,}
\author{P.F.Sofronov,}
\author{V.I.Stolyarsky.}

\address{Research Institute for Nuclear Problems,
Belarussian State University, 11~Bobruyskaya~Str. , Minsk 220050,
Belarus}

\begin{abstract}
Experimental studies of the backward wave oscillator with a
''grid'' diffraction grating that is formed by a periodic set of
metallic threads inside a rectangular waveguide are proceeded with
study of the generation intensity as a function of the grating
length.
\end{abstract}

\begin{keyword}
Volume Free Electron Laser (VFEL) \sep Volume Distributed Feedback (VDFB)
\sep diffraction grating \sep backward wave oscillator
\PACS 41.60.C \sep 41.75.F, H \sep 42.79.D
\end{keyword}

\end{frontmatter}

\qquad
Generation of radiation in millimeter and far-infrared
range with nonrelativistic and low-relativistic electron beams
gives rise difficulties. Gyrotrons and cyclotron resonance
facilities are used as sources in millimeter and sub-millimeter
range, but for their operation magnetic field about several tens
of kiloGauss ($\omega \sim \frac{eH}{mc}\gamma $) is necessary.
Slow-wave devices (TWT, BWT, orotrons)in this range require
application of dense and thin ($<0.1$ mm) electron beams, because
only electrons passing near the slowing structure at the distance
$\leq \lambda \beta \gamma /(4\pi )$ can interact with
electromagnetic wave effectively.
It is difficult to guide thin beams near slowing structure with
desired accuracy. And electrical endurance of resonator limits
radiation power and density of acceptable electron beam.
Conventional waveguide systems are essentially restricted by the
requirement for transverse dimensions of resonator, which should
not significantly exceed radiation wavelength. Otherwise,
generation efficiency decreases abruptly due to excitation of
plenty of modes. The most of the above problems can be overpassed
in VFEL
\cite{PhysLett,VFELreview,FirstLasing,FEL2002,patent}.
In VFEL the greater part of electron beam interacts with
the electromagnetic wave due to volume distributed interaction.
Transverse dimensions of VFEL resonator could significantly exceed
radiation wavelength $D \gg \lambda $. In addition, electron beam
and radiation power are distributed over the whole volume that is
beneficial for electrical endurance of the system.

The electrodynamical properties of volume diffraction structures
composed from strained dielectric threads was experimentally
studied in \cite{VolumeGrating}.
The electrodynamical properties of a "grid" volume resonator
formed by a perodic structure built from the metallic threads
inside a rectangular waveguide was considered in \cite{resonator}.
%
First observation of lasing of the backward wave oscillator with a
''grid'' diffraction grating and the volume FEL with a ''grid''
volume resonator that is formed by the periodic set of metallic
threads inside a rectangular waveguide  was described in
\cite{resonator-ex}.

In the present paper dependence of the generation intensity as a
function of the grating length is studied for the backward wave
oscillation regime.

\begin{figure}[h]
\epsfxsize = 12 cm \centerline{\epsfbox{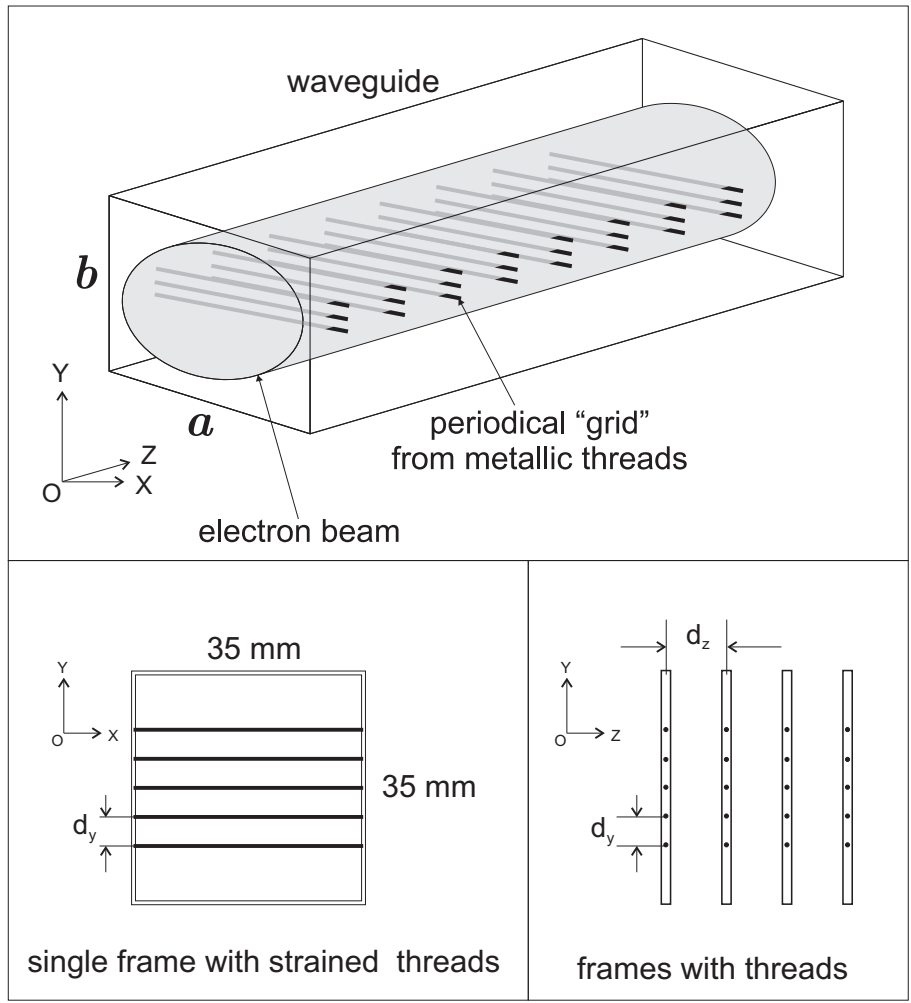}}
\caption{The  ''grid'' diffraction grating placed inside the
waveguide} \label{resonator}
\end{figure}

The  ''grid'' diffraction grating is built from tungsten threads
with diameter 0.1 mm strained inside the rectangular waveguide
with the cross-section 35 mm x 35 mm and length 300 mm (see
Fig.\ref{resonator}). The distance between the threads along the
axis $OZ$ is $d_z=12.5$ mm. A circular electron beam with the
diameter 32 mm, the energy $\sim$200 keV and current $\sim$2kA
passes through the above structure. Period of grating is chosen to
provide radiation frequency $\sim 8.4$ GHz.
The ''grid'' structure is made of separate frames each containing
the layer of 1, 3 or 5 parallel threads with the distance between
the next threads $d_y=6$ mm).
Joining frames provides to get the ''grid'' structure with layers
distant $d_z$ each from  other.

The purpose of the experiment is to study dependence of the
generated radiation intensity on the ''grid'' grating length (i.e.
on the number of frames).
Two types of experiments are reported: with one thread in the
frame and with 5 threads in the frame.

\begin{figure}[h]
\epsfxsize = 8 cm \centerline{\epsfbox{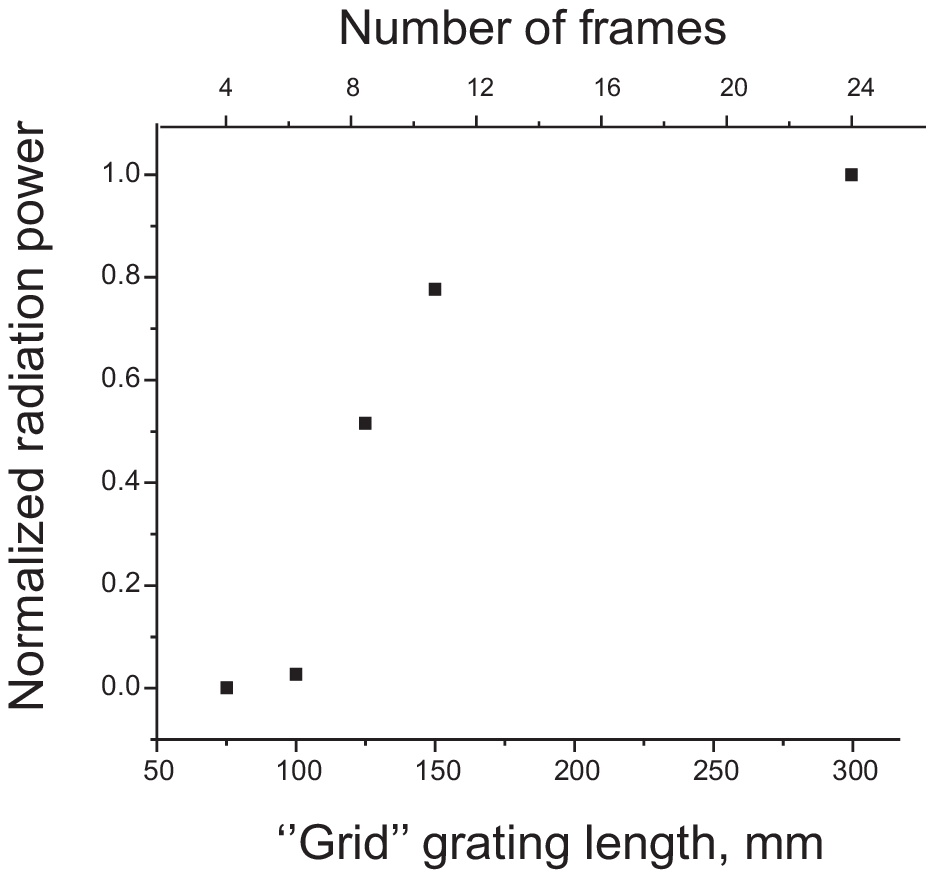}}
\caption{Dependence of the generation intensity on the ''grid''
grating length for 1 thread in the frame} \label{one}
\end{figure}

In the case of one thread in the frame the maximal radiation power
is about 1.5 kWatt. The radiation power is measured for 4, 8, 10
and 24 frames each containing one thread equidistant from
waveguide top and bottom walls. The result of these measurements
is presented in Fig.\ref{one}, where the radiation power is
normalized to the maximal detected power.

\begin{figure}[h]
\epsfxsize = 8 cm \centerline{\epsfbox{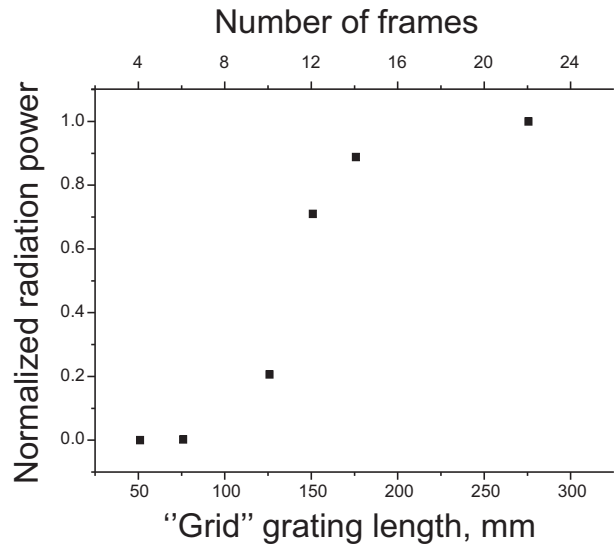}}
\caption{Dependence of the generation intensity on the ''grid''
grating length for 5 threads in the frame} \label{five}
\end{figure}

In the case of five threads in the frame the maximal radiation
power is about 8 kWatt. The radiation power is measured for 4, 6,
10, 12, 14 and 22 frames each containing five threads distant
$d_y=6$ mm each from other (see Fig.\ref{resonator}). The result
of these measurements is presented in Fig.\ref{five}, where the
radiation power is also normalized to the maximal detected power.

\begin{thebibliography}{}

\bibitem{PhysLett} V.G.~Baryshevsky, I.D.~Feranchuk,  {Phys.Lett.} {\bf 102A}, 141 (1984).

\bibitem{VFELreview} V.G.~Baryshevsky, {\ LANL e-print archive physics$/$9806039}.

\bibitem{FirstLasing}V.G.~Baryshevsky, K.G.~Batrakov, A.A.~Gurinovich et al.,
NIM {\bf  483A} (2002) 21-23.

\bibitem{FEL2002}V.G.~Baryshevsky, K.G.~Batrakov, A.A.~Gurinovich et al.,
NIM {\bf 507A} (2003) 137-140.

\bibitem{patent} Eurasian Patent no. 004665.

\bibitem{VolumeGrating} V.G.~Baryshevsky,K.G.~Batrakov,I.Ya.~Dubovskaya,V.A.~Karpovich,
V.M.~Rodionova, NIM {\bf 393A}, 71 (1997).
\bibitem{Nikolsky} V.V.~Nikolsky , Electrodynamics and propagation of radio-wave
(Nauka, 1978)

\bibitem{resonator} V.G.~Baryshevsky, A.A.~Gurinovich LANL e-print archive: physics/0409107

\bibitem{resonator-ex} V.G.~Baryshevsky, K.G.~Batrakov, N.A.~Belous, A.A.~Gurinovich, A.S.~Lobko, P.V.~Molchanov, P.F.~Sofronov,
V.I.~Stolyarsky, LANL e-print archive: physics/0409125.


\end{thebibliography}
\end{document}